\documentclass[11pt,a4paper]{article}
\usepackage{natbib}
\usepackage{lineno,hyperref}
\modulolinenumbers[5]
\pdfoutput=1

\usepackage{colortbl}
\usepackage{amsmath}
\usepackage{multirow}
\usepackage{setspace}
\usepackage[noend]{algpseudocode}
\usepackage[usenames,dvipsnames]{xcolor}
\usepackage{booktabs}
\usepackage{subcaption}
\usepackage{tikz}

\usepackage{lscape}
\usepackage{graphics}
\usepackage{graphicx}
\usepackage{caption}
\usepackage{csquotes}
\usepackage{longtable}
\usepackage{url}
\usepackage[shortlabels]{enumitem}

\usepackage{array}
\usepackage{rotating}
\usepackage[normalem]{ulem}
\usepackage[titletoc,title]{appendix}
\usepackage{bigstrut}
\usepackage[linesnumbered,ruled,vlined]{algorithm2e}
\usepackage{setspace}

\usepackage{amssymb}

\providecommand{\keywords}[1]
{
  \small	
  \textbf{\textit{Keywords---}} #1
}

\title{Integer Linear Programming for the Tutor Allocation Problem: A Practical Case in a British University}

\author{Giulia Caselli$^a$\footnote{Contact: Giulia Caselli. \textit{E-mail address:} giulia.caselli@unimore.it} , Maxence Delorme$^b$ , Manuel Iori$^a$}
\date{%
    $^a$\small\textit{DISMI, Universit{\` a} degli Studi di Modena e Reggio Emilia\\ Via Amendola 2, Pad. Morselli, 42122 Reggio Emilia, Italia}\\
    $^b$\textit{School of Mathematics, The University of Edinburgh, James Clerk Maxwell Building, The King's Buildings, Mayfield Road, Edinburgh EH9 3JZ, United Kingdom}\\[2ex]%
}

\begin{document}
\maketitle

\begin{abstract}
In the Tutor Allocation Problem, the objective is to assign a set of tutors to a set of workshops in order to maximize tutors' preferences. The problem is solved every year by many universities, each having its own specific set of constraints. In this work, we study the tutor allocation in the School of Mathematics at the University of Edinburgh, and solve it with an integer linear programming model. We tested the model on the 2019/2020 case, obtaining a significant improvement with respect to the manual assignment in use. Further tests on randomly created instances show that the model can be used to address cases of broad interest. We also provide meaningful insights on how input parameters, such as the number of workshop locations and the length of the tutors' preference list, might affect the performance of the model and the average number of preferences satisfied.
\end{abstract}


\section{Introduction}
\label{sec:Intro_TAP}
As in many other countries, British academic courses are divided into lectures and workshops. 
While lectures bring the necessary theoretical background to the students, workshops allow them to put their newly acquired knowledge in practice. 
Lectures are delivered by {\em lecturers} and workshops are delivered by {\em tutors}.
In most courses, only one lecturer is responsible for the lectures during the academic year.
As most lecturers have open-ended contracts with their universities, once a lecturer is assigned to a course, there is an interest in keeping it over the years: the preparation time becomes shorter and the teaching abilities increase. 
Such consistency is unfortunately not possible for the workshops: indeed, as the majority of the tutoring staff is composed of PhD students and post-doctoral researchers, it changes every year. 
In addition, as workshops require more than one tutor (depending on the ratio tutors/students wished by the university), the problem of assigning tutors to workshops, also called the {\em Tutor Allocation Problem} (TAP), is far from trivial and has to be solved every year, representing a significant workload for the administrative team. 

In the School of Mathematics at the University of Edinburgh, the TAP is solved manually by the administrative staff in August, before the beginning of every academic year. 
An update taking into account the staff changes is also performed in November. 
Even though the TAP is solved manually, the decision makers have to take into account several parameters. 
First, tutors have different sets of skills, and high quality workshops require tutors with the appropriate skills. 
Second, different kinds of constraints have to be respected: for example, some tutors have a maximum number of tutoring hours specified in their working contracts and there are incompatibility constraints between two simultaneous workshops. Third, some tutors have a preference list (not ranked in our case), that consists of a subset of courses that they wish to tutor. In general, a tutor is happier and gives better tutoring if she is assigned to a workshop in her preference list. Another key aspect of the TAP is the fact that the university is a dynamic environment: indeed, the members of staff may change unexpectedly, or the number of students in a given workshop may increase, requiring an additional tutor. The slightest update in the data makes the assignment obsolete and requires a new solution. 

Currently, the School of Mathematics first assigns lecturers to lectures, then schedules the lectures and workshops (together with the room allocation), and then uses an iterative process to solve the TAP. Based on the tutors' preferences, skills, and availabilities, an initial assignment is proposed to the faculty members. 
After collecting the requests for changes, another solution with minor modifications is proposed. The process may be iterated a few times before the final assignment is reached. 

In this work, our goal is to ease the work by the administrative staff by proposing an {\em Integer Linear Programming} (ILP) model whose goal is to provide the initial assignment. The ILP model has several advantages. First, it ensures that all the constraints are satisfied (we observed that a manual assignment sometimes allocates a tutor to two simultaneous workshops, resulting in a request for changes). Second, as our objective function is to maximize the tutors' satisfaction, it provides better allocations for the tutors. Third, it can be easily generalized to include further constraints, thus making it a powerful tool for addressing several real-world situations.

The rest of the paper is organized as follows. A concise description of the related literature is given in Section~\ref{sec:rev_TAP}. {In Section \ref{sec:problem_TAP}, we provide a detailed description of the problem. In Section \ref{sec:model_TAP}, we formally present our mathematical model.} Section \ref{sec:exp_TAP} reports the computational experiments, including the main test on the real-world case study we face, as well as a number of additional tests performed on randomly created instances with the aim of assessing the model behavior and evaluating the impact of some key input parameters. Concluding remarks are finally provided in Section \ref{sec:conclusion_TAP}.

\section{Concise literature review} \label{sec:rev_TAP}
The TAP and its related problems have been studied since the early seventies (see \cite{AC71}). 
We refer the reader to the survey of \cite{CL98}, who reviewed in details the papers related to real-world allocation problems faced by universities. In the following, we mention some of the most important mathematical models and heuristics proposed in the literature for the TAP and closely related problems. To ease the comprehension, we always use the TAP terminology (i.e., tutors and workshops) to describe the models. We also provide a recapitulative table summarizing the main features of each problem at the end of the section (Table \ref{tab:lit}).

In the late seventies, \cite{DM76} proposed an ILP model to assign workshops to tutors. The model maximizes the tutors' preferences, while taking into account a minimum/maximum number of tutors per workshop, and a minimum/maximum number of workshops per tutor. The model also takes into account a large time period (such as a quarter or a semester) in order to balance the tutors' workload. They showed that the problem could be transformed into a network flow problem, but did not provide computational experiments. A few years later, a similar model was used by \cite{HNC83} for solving a more general class of problems. Their model was tested on two real-world problems: assigning students to job interview slots and assigning students to project teams in an MBA field project course. The model could also take into account the tutors' skills and make sure that a minimum number of tutors with the appropriate skills were assigned to the workshops. \cite{DMV89} also proposed a network flow formulation for an extension of the TAP in which the room and the schedule is determined by the model.

As noticeable variations of the previous models, we mention the work of \cite{B76}, who proposed a model based on continuous variables used to assign percentages of workshops to tutors. The model still maximizes the tutors' preferences, however, it also takes into account the notion of time slots, making sure that a tutor is not assigned to two simultaneous workshops. We also mention the work of \cite{T75}, who introduced a zero-one programming model in which the objective function combines tutors' preferences and effectiveness. In the work of \cite{YP89}, the tutors' effectiveness is measured through 15 different scores performed by a survey given to the students. \cite{HL75}, \cite{LS83}, and \cite{AHA11} also used an advanced objective function involving positive and negative deviations from the tutors' targeted numbers of hours. 

\cite{SS77} presented a two-stage optimization approach: in the first stage, tutors are assigned to workshops maximizing their preferences; in the second stage, workshops are assigned to time slots. 
\cite{HC97} proposed a model for the TAP that aimed at minimizing the number of different courses the tutors are assigned to. In other words, the model prefers assigning a tutor to many workshops of few courses  rather than few workshops of many courses. The authors also pointed out the resemblance between their version of the TAP and the fixed charge transportation problem. 

\cite{MW84} introduced the first model in which tutors are assigned to a pattern of workshops. As every pattern respects the tutoring load, the model has very few constraints. However, it has a very large number of variables, as the number of tutoring patterns is exponential. They later extended their work in \cite{MW87} to allow a more advanced objective function involving positive and negative deviations from the tutors' targeted numbers of hours. Theoretical investigations on such models were provided in \cite{BQB11}. 

An ILP model for the TAP with ad hoc constraints was proposed by \cite{AS06}. Their model considers incompatibility constraints for simultaneous and successive workshops, and gender-based policies. Another model for the TAP, specifically targeted for tutors and workshops, was proposed by \cite{AS17}. Due to the exponential number of variables the latter model involved, they proposed a column generation procedure. {Over the last two decades, many other mathematical and heuristic approaches have been proposed, each of them involving specific constraints related to the authors' institution. We mention in particular the work of \cite{W02} in Taiwan,  \cite{GNO08} in Indonesia, \cite{GKDA15} in Turkey, \cite{DD18} in Brazil, and \cite{DL16} in Spain.

Very recently, other problems closely related to the TAP have been studied. \cite{GMR17} addressed the issue of remedial education for underprepared students that need additional support by teachers. They focused on high school remedial education, although they explicitly underlined the problem applicability to university systems. Assignment of teachers to courses and timetabling were modeled into a unique ILP formulation, and a heuristic approach was implemented to solve large size instances.

The aim of our work is to provide an ILP model able to solve excatly the TAP with constraints that are specific to our case study. Table \ref{tab:lit} gives a summary of the TAP models previously mentioned, with indicators on whether or not they consider specific features. The last line is specific to our case. The following notation is used in the table: 
\begin{itemize} 
    \item \emph{Nature of the objective function}: O1: Multi-objective function (goal programming); O2: Balanced function (weight factors); O3: Utility function (based on tutors' preferences or productivity);
    \item \emph{Sets of constraints}: C1: Workshops need a minimum/exact number of tutors (or tutoring hours); C2: Tutors can be assigned to a minimum and/or maximum number of workshops (and/or tutoring hours); C3: Tutors can be assigned to maximum one (pre-determined) configuration of workshops for each course; C4: Tutors are eligible for workshops (for instance in terms of knowledge, computer skills, academic position and so on); C5: Tutors cannot be assigned to two concurrent workshops; C6: Tutors cannot be assigned to two following workshops located in different sites if they don't have the time required for travel; C7: Some tutors-workshops assignments are forced by the model; C8: Some tutors-workshops assignments are forbidden by the model.
 \item \emph{Domain of the decision variables}: V1: Binary decision variables; V2: Integer decision variables; V3: Non negative continuous decision variables.
\end{itemize}

\begin{table}[bt]
	\centering
	\caption{Literature review of the main models for the personnel assignment problem} \label{tab:lit}
	\footnotesize
	\setlength{\tabcolsep}{0.125cm}
	\resizebox{\textwidth}{!}{%
	\begin{tabular}{lcccccccccccccc}
		\toprule 
 		\multicolumn{1}{l}{{Model}} & \multicolumn{3}{c}{{Obj. func.}} &
		\multicolumn{8}{c}{{Constraints}} & \multicolumn{3}{c}{{Variables}}\\ 
		 \cmidrule(lr){2-4}  \cmidrule(lr){5-12}  \cmidrule(lr){13-15}
		\multicolumn{1}{c}{} & \multicolumn{1}{c}{{O1}} & \multicolumn{1}{c}{{O2}} & \multicolumn{1}{c}{{O3}} &
		\multicolumn{1}{c}{{C1}} & \multicolumn{1}{c}{{C2}} & \multicolumn{1}{c}{{C3}} & \multicolumn{1}{c}{{C4}} & \multicolumn{1}{c}{{C5}} & 		\multicolumn{1}{c}{{C6}} & \multicolumn{1}{c}{{C7}} & \multicolumn{1}{c}{{C8}} & \multicolumn{1}{c}{{V1}} & \multicolumn{1}{c}{{V2}} & 		\multicolumn{1}{c}{{V3}}  \\
		\cmidrule(lr){1-1} \cmidrule(lr){2-4}  \cmidrule(lr){5-12}  \cmidrule(lr){13-15}
		\cite{AHA11} & \checkmark &  & \checkmark & \checkmark & \checkmark &  &  &  &  &  &  & \checkmark &  & \checkmark  \\ 
		\cite{AS06} &  & \checkmark & \checkmark & \checkmark & \checkmark	&  &  & \checkmark &  &  &  & \checkmark &  &   \\ 
		\cite{AS17} &  &  & & \checkmark & \checkmark &  & \checkmark & \checkmark & \checkmark &  &  & \checkmark & \checkmark &   \\  
		\cite{B76}  &  &  & \checkmark & \checkmark & \checkmark &  &  & \checkmark &  &  &  &  &  & \checkmark \\
		\cite{BQB11} &  &  &  & \checkmark & & \checkmark &  &  &  &  &  & \checkmark &  &  \\
		\cite{DD18}  &  & \checkmark & & \checkmark & \checkmark &  &  &  &  &  &  & \checkmark &  & \checkmark  \\
		\cite{DMV89} & \checkmark &  & \checkmark & \checkmark & \checkmark	&  &  & \checkmark &  &  &  &  & & \checkmark \\ 
		\cite{DL16} &  & \checkmark & \checkmark & \checkmark & \checkmark &  &  & \checkmark &  &  &  & \checkmark &  & \checkmark \\
		\cite{DM76} &  &  & \checkmark & \checkmark & \checkmark &  &  &  &  &  &  &  & \checkmark &  \\
		\cite{GMR17} &  & \checkmark &  & \checkmark &  &  & \checkmark & \checkmark &  &  &  & \checkmark  &  &  \\	
		\cite{GKDA15} & \checkmark &  & \checkmark & \checkmark & \checkmark &  &  & \checkmark &  &  &  & \checkmark &  & \checkmark  \\
		\cite{GNO08} &  & \checkmark & & \checkmark & \checkmark &  &  &  &  & \checkmark &  & \checkmark &  &   \\
		\cite{HL75} & \checkmark &  & &  & \checkmark &  &  & \checkmark &  &  &  & \checkmark &  & \checkmark  \\
		\cite{HNC83} &  &  & \checkmark & \checkmark & \checkmark &  & \checkmark &  &  &  &   & \checkmark &  & \\ 
		\cite{HC97} &  &  & &\checkmark & \checkmark &  &  &  &  &  &  & \checkmark  & & \checkmark \\
		\cite{LS83} & \checkmark &  & \checkmark &  &  &  &  &  &  &  &  & \checkmark &  & \checkmark  \\
		\cite{MW84} &  &  & \checkmark & \checkmark &  & \checkmark &  &  &  &  &  & \checkmark &  &  \\
		\cite{MW87} & \checkmark &  & \checkmark & \checkmark &  & \checkmark &  &  &  &  &  & \checkmark &  &  \\
		\cite{SS77} &  &  & \checkmark & \checkmark & \checkmark &  &  &  &  &  &  & \checkmark &   &   \\
		\cite{T75} &  & \checkmark & \checkmark & \checkmark & \checkmark & \checkmark &  &  &  &  &  & \checkmark &  &   \\
		\cite{W02} & & & \checkmark & \checkmark & \checkmark & & \checkmark & & & & & \checkmark & & \\ 
		\cite{YP89} &  &  & \checkmark & \checkmark & \checkmark	&  &  &  &  &  &  & \checkmark &  &   \\ 
		\cmidrule(lr){1-1} \cmidrule(lr){2-4}  \cmidrule(lr){5-12}  \cmidrule(lr){13-15}
		Our case  &  &  & \checkmark & \checkmark & \checkmark & \checkmark & \checkmark & \checkmark & \checkmark & \checkmark  & \checkmark & \checkmark &  &  \\
		\bottomrule
	\end{tabular}
	}
\end{table}

\section{Problem description} \label{sec:problem_TAP}
{In this section, we describe each component of the TAP and give an exhaustive list of the constraints that need to be considered. }

{\subsection{Definitions}}\label{subsec:pb_TAP}
\paragraph*{Courses}
A course is a combination of lectures and workshops. Each course has a set of research groups to which it is associated and is delivered at a specific academic year (e.g., first year, second year, post-graduate). On the students' side, a course is composed of at most one lecture and one workshop per week. As the amphitheatres are large enough to accommodate hundreds of students, lectures are given only once. However, due to certain restriction such as the room size or the high number of participants, it is possible that the same workshop has to be repeated several times during the week. We identified three types of courses: small courses, with one lecture and one workshop section per week; medium courses, with one lecture and two or three workshop sections per week, generally in different days of the week; and large courses, one lecture and four or more consecutive workshop sections in the same day of the week.     

As far as scheduling is concerned, some courses are proposed twice during the year (once per semester) while some others are semester-specific. Lectures and workshops are delivered either every week, every even week, or every odd week. 

\paragraph*{Lectures}
Each lecture is given by one lecturer and has a location, a day, and a time slot. All these data are fixed (i.e., determined before solving the TAP), so we consider them as inputs. They are used to define scheduling incompatibilities. For example, a member of staff cannot lecture a course and tutor another at the same time.

\paragraph*{Workshops}
Each workshop has a {\em supertutor}, a location, a day, a time slot, and a set of computer skills (e.g., softwares Maple or Matlab) that are fixed, and a set of tutors that needs to be determined.

\paragraph*{Supertutors}
Each workshop section has a supertutor that is fixed. The main task of the supertutor is to manage the other tutors, and possibly answer the questions tutors are not able to answer. In most cases, the lecturer of a given course is also the supertutor for the workshops of that course. Supertutoring hours are included in the tutoring hours.

\paragraph*{Tutors}
Tutors refer to all the faculty members of the department (academics, postdocs, PhD students) and external members (hired as consultants or members of other departments). Tutoring is mandatory for some of the academic members (e.g., when the working contracts involve a minimum number of tutoring hours to do), while it is based on volunteering for some others (e.g., some postdocs wish to tutor to get some teaching experience).

A minimum and a maximum number of hours of tutoring are assigned to each tutor every semester. This number depends on the category in which the tutor belongs and her specific requests (e.g., some members prefer to do all their tutoring hours in one semester).
In addition, the tutors have a maximum number of courses they are allowed to be involved with each semester to prevent someone to be assigned to many different small courses. For a standard tutor, with a balanced workload between semesters, the limit is set to three different courses per semester.

Each tutor belongs to one or several research groups, may have one or several computer skills, and is allowed to tutor courses from a specific set of years. In addition, tutors have a preference list of courses they wish to tutor. The preferences are expressed through the annual tutor survey, together with all the special requests in terms of availability, including sabbatical semesters and extra commitments. A tutor can only be assigned to a workshop if she has all the appropriate skills, if she has at least one of the appropriate research group, and if she is allowed to tutor at the level (year) of the course.

\subsection{System requirements}
\label{subsec:req_TAP}
The specific requirements of the system described in this problem are the following: 
\begin{enumerate}
    \item a tutor can be assigned to a workshop only if her research group, the years she is allowed to tutor, and her skills are consistent with those required by the workshop. Note that, as the skills depend on the workshop and not on the course, a tutor may be assigned to only certain workshops of a given course;
    \item the number of required tutors for each workshop section must be respected;
    \item the minimum and maximum number of hours and courses per semester for each tutor must be respected; the number of hours spent in a given workshop is the multiplication between the actual number of hours spent in that workshop and a parameter called the {\em Tutor Marking Multiplier (TMM)}, which is assigned to each course as a scaling factor so that certain activities are taken into account (e.g., marking);
    \item each tutor can be assigned to multiple tutoring sections of the same course, but no more than three in a row;
    \item workshops are delivered in two locations, i.e., two campuses; each daily movement from one campus to the other requires at least one hour to travel;
    \item an academic should not be supertutor and tutor of the same course, even on different workshops;
    \item the allocation should satisfy tutors' preferences as much as possible.
\end{enumerate}

The administrative staff in charge of the tutor allocation is also able to give additional preferences based on historical data, remove  preferences they believe are not adequate, or force/forbid a tutor to be assigned to a given course.

\subsection{Model setting} \label{subsec:setting_TAP}
In order to clarify the model formulation reported in the following section, we introduce the concept of {\em configuration} and {\em incompatibilities}.
\paragraph*{Configurations}
A configuration is a subset of workshop sections scheduled for one specific course that respect the system requirements: it can be a single section, two or three following sections in the same day, or multiple sections in different days. Feasible configurations are determined in a separate procedure called before the model. 
We say that a configuration is {\em active} at a given time slot if one workshop included in the configuration occurs at this time slot. 
Each tutor can be assigned to at most one configuration of each course. If a tutor cannot be assigned to a given course, then we prevent her to be assigned to any configuration of that course.
\paragraph*{Incompatibilities}
We identified two types of incompatibilities: (i) for each time slot, no tutor can be assigned to more than one configuration that is active in that time slot; (ii) a tutor cannot be assigned to two configurations that are active in consecutive time slots if they are in two different locations.
\paragraph*{Forced and forbidden configurations}
Some specific system requirements have to be included in the model as forced configurations. For instance, externals not working in the department are included in the list of tutors because they are needed for a specific course. Therefore, we have to make sure that they are assigned to the appropriate configuration of the course they are hired for. Similarly, we use forbidden configurations to prevent them to be assigned to any configuration of a course they were not hired to tutor. We also forbid all the configurations a tutor cannot do, either because she is lecturing during one of the time slot in which the configuration is active, or because she does not have the appropriate skills, research group, or year.

\section{Mathematical model} \label{sec:model_TAP}
In this section, an ILP model is formulated to address the TAP. The needed mathematical notation is defined in Table \ref{tab:param}.
\begin{table}[hbt]
	\centering
	\caption{Mathematical notation}
	\label{tab:param}
	\begin{tabular}{l l}
		\toprule
		Notation & Definition \\
		\hline
		$I$ & Set of available tutors in the semester \\
		$K$ & Set of courses scheduled in the semester \\
		$S$ & Set of courses' sections requiring tutors  \\
		$J$ & Set of courses' configurations to be assigned to tutors \\
		$J_K$ & Set of possible configurations for course $k$ \\
		$J_S$ & Set of configurations including section $s$ \\
		$D$ & Set of teaching days in the considered semester ($D = 1,2, \dots$) \\
		$T$ & Set of time slots in a teaching day ($T = 1,2,\dots$)\\
		$P_i$ & Set of preferences  for tutor $i$ (configurations of preferred courses) \\
		$F_i$ & Set of forced configurations which tutor $i$ is forced to tutor \\
		$F'_i$ & Set of forbidden configurations that cannot be tutored by tutor $i$ \\
		$l_i, u_i$ & Min/max number of tutoring hours per semester for tutor $i \in I$ \\
		$L_i, U_i$ & Min/max number of allocated courses per semester for tutor $i \in I$ \\
		$h_j$ & Total number of tutoring hours per semester for configuration $j \in J$ \\
		$t_j$ & TMM of course which configuration $j \in J$ belongs to \\
		$N_s$ & Number of tutors required by each course section $s$ \\
		$m_i$ & Number of preferences (courses) expressed by tutor $i$\\
		\bottomrule
	\end{tabular}
\end{table}

We introduce a set of two-index binary variables $x_{ij}$ that take the value 1 if tutor $i$ is assigned to configuration $j$, and 0 otherwise ($i \in I, j \in J)$. 
We also set $T' = T \setminus\{T_{\max}\}$ and $T'' = T \setminus\{\{T_{\max}\} \cup \{T_{\max}-1\}\}$, and we define
$C(d, t, l)$ as the set of all configurations $j$ that are active at time slot $t$ of day $d$ at location $l$ ($j \in J, d \in D, t \in T, l \in \{1,2\}$).
The TAP is modelled as follows:
	\allowdisplaybreaks
\begin{align}
	\label{eq1} & \max &&\sum_{i \in I}\sum_{j \in J} \frac{1}{\min\{m_i,U_i\}} x_{ij}\\
	\label{eq2}&\text{s.t.} &&\sum_{i \in I}\sum_{j \in J_s} x_{ij} = N_{s}  \quad  &  s \in S \\
	\label{eq3} && l_i \le &\sum_{j \in J} h_jt_{j}x_{ij} \le u_i  \quad  &  i \in I \\
	\label{eq4} &&  L_i \le &\sum_{j \in J} x_{ij} \le U_i  \quad  &  i \in I \\
	\label{eq5}	&&& \sum_{j \in Jk} x_{ij} \le 1  \quad  & i \in I  \quad  & k \in K \\
	\label{eq9} &&&  \sum_{l \in \{1,2\}} \sum_{j \in C(d,t,l)}x_{ij} \le 1 \quad  & i \in I, \quad   d \in D, \quad  & t \in T \\
	\label{eq10} &&&  \sum_{j \in C(d,t,1)}x_{ij} + \sum_{j \in C(d,t+1,2)}x_{ij}  \le 1 \quad  & i \in I,  \quad  d \in D, \quad  & t \in T'\\
	\label{eq11} &&&  \sum_{j \in C(d,t,2)}x_{ij} + \sum_{j \in C(d,t+1,1)}x_{ij}  \le 1 \quad  & i \in I, \quad  d \in D, \quad  & t \in T'\\
	\label{eq12} &&&  \sum_{j \in C(d,t,1)}x_{ij} + \sum_{j \in C(d,t+2,2)}x_{ij}  \le 1 \quad  & i \in I,  \quad  d \in D, \quad  & t \in T'' \\
	\label{eq13} &&&  \sum_{j \in C(d,t,2)}x_{ij} + \sum_{j \in C(d,t+2,1)}x_{ij}  \le 1 \quad & i \in I, \quad  d \in D, \quad  & t \in T'' \\
	\label{eq14}	&&& x_{ij} = 1  \quad  & i \in I \quad  & j \in F \\
	\label{eq15}	&&& x_{ij} = 0  \quad  & i \in I \quad  & j \in F' \\
	\label{eq16}	&&&  x_{ij} \in \{0,1\}  \quad  & i \in I \quad & j \in J
\end{align}
The objective function (\ref{eq1}) maximizes tutors' preferences. Tutors are free to express as many preferences as they want in the annual survey and their preference lists may be shortened or made longer by the administrative staff. 
Coefficients $\frac{1}{\min\{m_i,U_i\}}$ make sure that every tutor has the same weight, whether they listed a number of preferences greater than, equal to, or lower than $U_i$.
Constraints (\ref{eq2}) ensure that the required number of tutors is allocated to each workshop section. Constraints (\ref{eq3}) and (\ref{eq4}) ensure that the number of tutoring hours and courses  allocated to each tutor respect the given limits.
Constraints (\ref{eq5}) are necessary to guarantee that each tutor is assigned to at most one configuration for each course. 
Constraints (\ref{eq9}) represent incompatibility constraints for concurrent sections: each tutor can be at most in one location in each time slot. 
Constraints (\ref{eq10})-(\ref{eq13}) are location incompatibility constraints. Courses are distributed in two locations in the presented model and a fixed number of time slots (2 in our case) is required for tutors to move from one location to another so they cannot be assigned to any activity during those time slots. 
Constraints (\ref{eq14}) and (\ref{eq15}) refer to forced and forbidden configurations respectively. Note that the preprocessing of most mathematical solvers automatically removes the variables set to 0 in (\ref{eq15}).
Finally, constraints (\ref{eq16}) state that variables $x_{ij}$ are binary.

For the sake of conciseness, the model only considers one semester, but can easily be extended to handle both semesters simultaneously by making a copy of constraints \eqref{eq3} and \eqref{eq4} per semester. Constraints related to the skills, the research group, and the academic years are included in the forbidden configurations constraints \eqref{eq15}. {These three aspects are related to tutors' teaching competences and play a key role in our model to provide a valuable assignment solution. Indeed, research puts a significant effort in the study of competence analytics and good results have been obtained in the literature with the use of similar mathematical approaches, as shown by, e.g., \cite{BMKTAF17}.}

\section{Computational experiments} \label{sec:exp_TAP}
In this section, we study the performances of model \eqref{eq1}-\eqref{eq16} on a real TAP instance (from the University of Edinburgh, School of Mathematics, academic year 2019-2020), and on randomly generated instances with larger sizes. Our model has been coded in C++ and solved with Gurobi 8.1.1. on an Intel Core i7, 1.80 GHz, with 16 GB of RAM memory, running under Windows 10 64 bits. A time limit of 3600 seconds per instance has been imposed to the solver.

\subsection{Case study} \label{subsec:casestudy_TAP}
We ran our model on two semesters, for a total of 115 days (11 weeks in the first semester and 12 weeks in the second semester). 
Activities are scheduled from Monday to Friday, between 9 am and 6 pm, so there are 18 time slots of 30 minutes every day. 
Lectures and workshops of the School of Mathematics are distributed in two campuses, one hour distant from each other (around 95\% of the workshops take place in the first location, 5\% in the second). Around one hundred courses are proposed in the 2019-2020 academic year, equally split among the two semesters. 
The TMM is usually equal to 2, but can take values between 1 and 2.5.
Almost 300 tutors are available for the considered academic year. 
The maximum number of assignable courses per semester is equal to 3 for most tutors with rare exceptions. 
The maximum number of tutoring hours per semester varies in general between 80 hours and 120 hours. 
There are 9 research groups in the School of Mathematics and each tutor belongs to one group or more. 
Courses are typically compatible with more than one research group. 
On average, a tutor is compatible with 37\% of the courses and expresses 1.4 preferences.

An optimal solution was found by the model within one minute. On average, tutors obtained 60\% of their preferences, which is an improvement with respect to the manual solution obtained by the administrative team that only gave them 31\% of their preferences. Note that only tutors who expressed at least one preference are considered when performing the indicator on proportion of satisfied preferences. Figure \ref{fig:realcase} shows a detailed comparison between the two solutions. Tutors are classified with respect to the number of preferences they expressed in the horizontal axis. For each group, the graph shows in the $y$ axis the total number of preferences expressed by the tutors, the number of preferences satisfied by the solution of our model, and the number of preferences satisfied by the manual assignment. We observe an interesting pattern: 
\begin{itemize}
    \item for the groups of tutors with few preferences (1 and 2), the manual assignment is slightly worse than our model;
    \item for the groups of tutors with a large number of preferences (3 and 4), the solution given by the manual assignment is significantly worse than our model;
    \item for the groups of tutors with a very large number of preferences (5 and 6), the solution given by the manual assignment is still worse than our model, but at a lesser extent.
\end{itemize}
One possible explanation is that the administrative team tries to satisfy as much as possible the tutors expressing 1 or 2 preferences. For tutors expressing 3 or 4 preferences, the administrative team tries to satisfy at least one of their preferences, if it is possible. Tutors with 5 preferences or more seem to have been in the system for several years, and it may be possible that the administrative staff tries to give a higher priority to these ``experienced" tutors. 

\begin{figure}[h]
    \centering
    \includegraphics[width=0.7\textwidth]{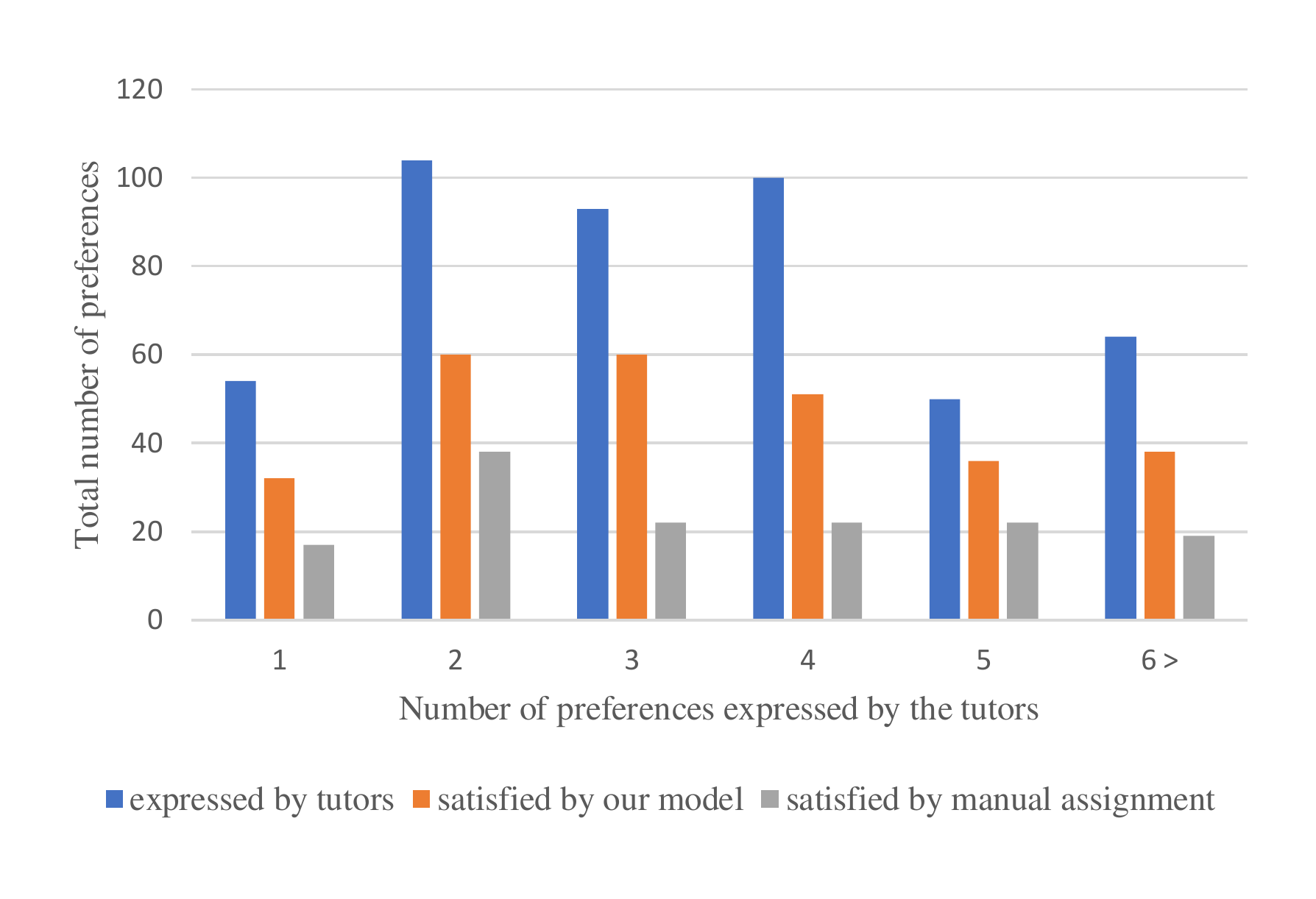}
    \caption{Total sum of preferences satisfied by the model solution and the manual assignment.}
    \label{fig:realcase}
\end{figure}
When we compared our solution with the manual solution, we observed that only a third of the allocations were similar, indicating that there is a high interest in using an ILP model to generate the initial solution of tutors to workshops, as this can lead to solutions that are very difficult to obtain manually.

\subsection{Experiment with random instances} \label{subsec:genexp_TAP}
To gain some deeper insight on the performance and scalability of the mathematical model, we also used it to solve randomly generated instances. We designed an ad-hoc data-generator that could reproduce the structure of our real-world instance, but in which we could also vary the number of tutors and courses. With the generator, we created 50 instances of 5 different sizes (number of tutors, number of courses), namely: (250, 100), (500, 200), (750, 300), (1000, 400), and (1500, 600). 

A summary of the average results we obtained is reported in Table \ref{tab:results} and the resulting behaviors are illustrated by the graphical representations in Figures \ref{fig:averageres} and \ref{fig:optimalsol}. In the table, apart from the numbers of tutors, courses and instances, we also report the numbers of feasible solutions found and proven optimal solutions found, as well as the average model execution time (in seconds) and the average solution value (computed as in \eqref{eq1}).

\begin{table}[bth]
	\caption{Results obtained with the ILP model}
	\label{tab:results}
	\footnotesize
	\centering
	\begin{tabular}{rrrrrrr}
		\toprule
		\multicolumn{1}{c}{N. tutors} & \multicolumn{1}{c}{N. courses} & \multicolumn{1}{c}{N. instances} & \multicolumn{1}{c}{N. feasible}  & \multicolumn{1}{c}{N. optimal}  & \multicolumn{1}{c}{AVG model} & \multicolumn{1}{c}{AVG solution} \\
			 & 	 & 	 &  \multicolumn{1}{c}{solutions} & \multicolumn{1}{c}{solutions} &  \multicolumn{1}{c}{time [s]} &   \multicolumn{1}{c}{value} \\
		\cmidrule(lr){1-3} \cmidrule(lr){4-7}
		250 & 100 & 50 & 46 & 46 & 56.27 & 102.01 \\
		500 & 200 & 50 & 47 & 47 & 186.50 & 205.05 \\
		750 & 300 & 50 & 40 & 40 & 397.15 & 303.67 \\
		1000 & 400 & 50 & 39 & 39 & 663.83 & 408.36 \\
		1500 & 600 & 50 & 28 & 28 & 1474.68 & 614.30 \\
		\bottomrule
	\end{tabular}
\end{table}

We observe that:
some instances are infeasible, meaning that the problem can be over-constrained sometimes;
all instances were solved, some of them with a proof of infeasibility, and the remaining ones with a proven optimal solution;
the model can cope even with large size instances (up to 1500 tutors and 600 courses);
on average, tutors obtain around 50\% of their preferences, which is consistent with our real instance.

Figure \ref{fig:averageres} shows the results in terms of average performance of experiments. The number of tutors is shown in the horizontal axis. The average calculation time and the average solution value (when available) are shown in the vertical axis. We observed that the solution value grows linearly with the number of tutors, indicating that, with the considered parameters, we can only satisfy around 50\% of the tutors' preferences. We also observe that the average running time does not grow linearly with the number of tutors, indicating that instances with 3000 tutors or more would be probably intractable by our model. However, it is quite unlikely that instances of such size appear in a real-world academic context.

\begin{figure}[ht]
	\centering
	\begin{subfigure}[t]{0.7\textwidth}
		\centering
		\includegraphics[width=1\textwidth]{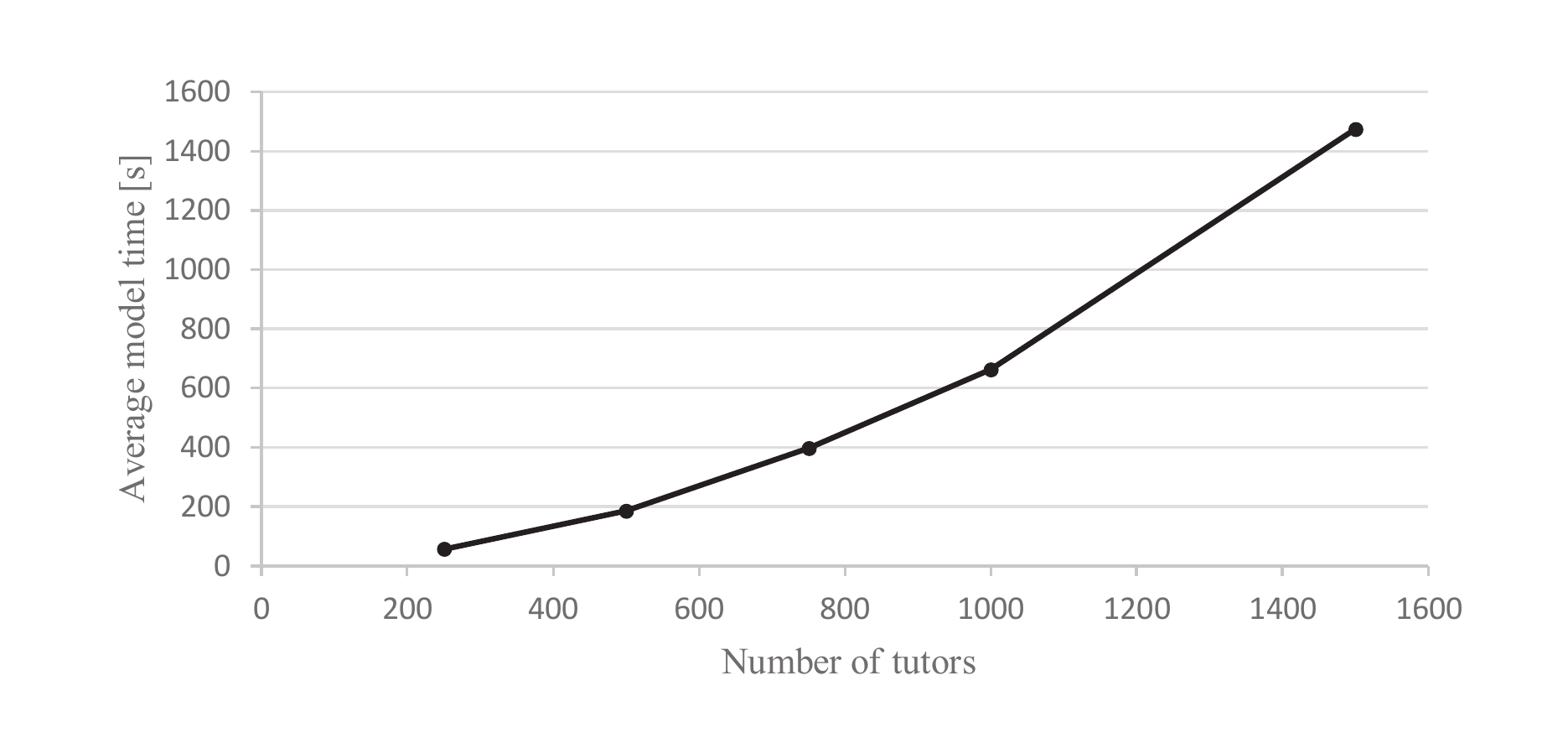}
		\caption{Average model calculation time}
	\end{subfigure}\hfill
	\begin{subfigure}[t]{0.7\textwidth}
		\centering
		\includegraphics[width=1\textwidth]{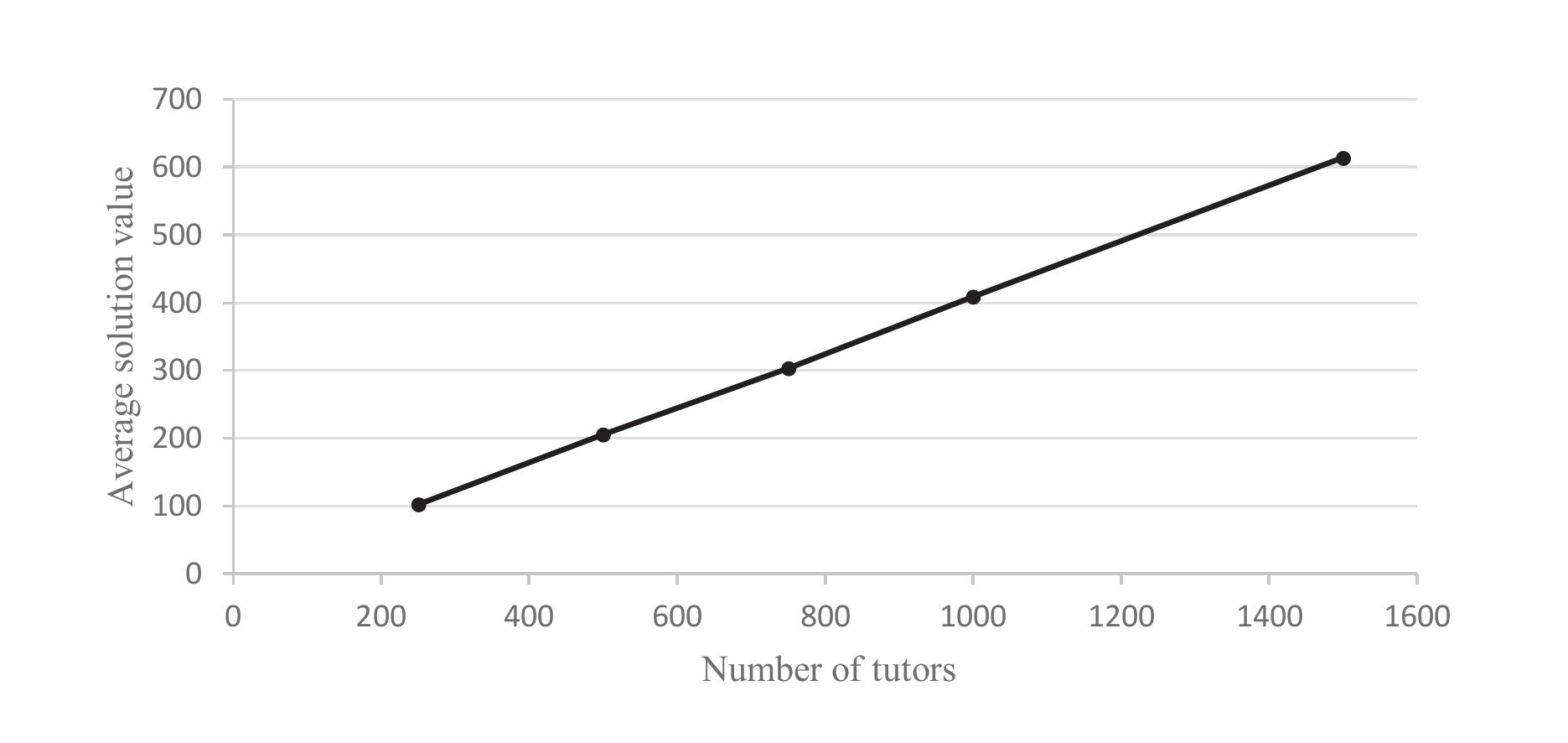}
		\caption{Average model solution value}
	\end{subfigure}
	\caption{Average model performance depending on the number of tutors (and courses)}\label{fig:averageres}
\end{figure}

 Figure \ref{fig:optimalsol} shows the percentage of feasible instances in the five experiments. 
\begin{figure}[ht]
    \centering
    \includegraphics[width=0.7\textwidth]{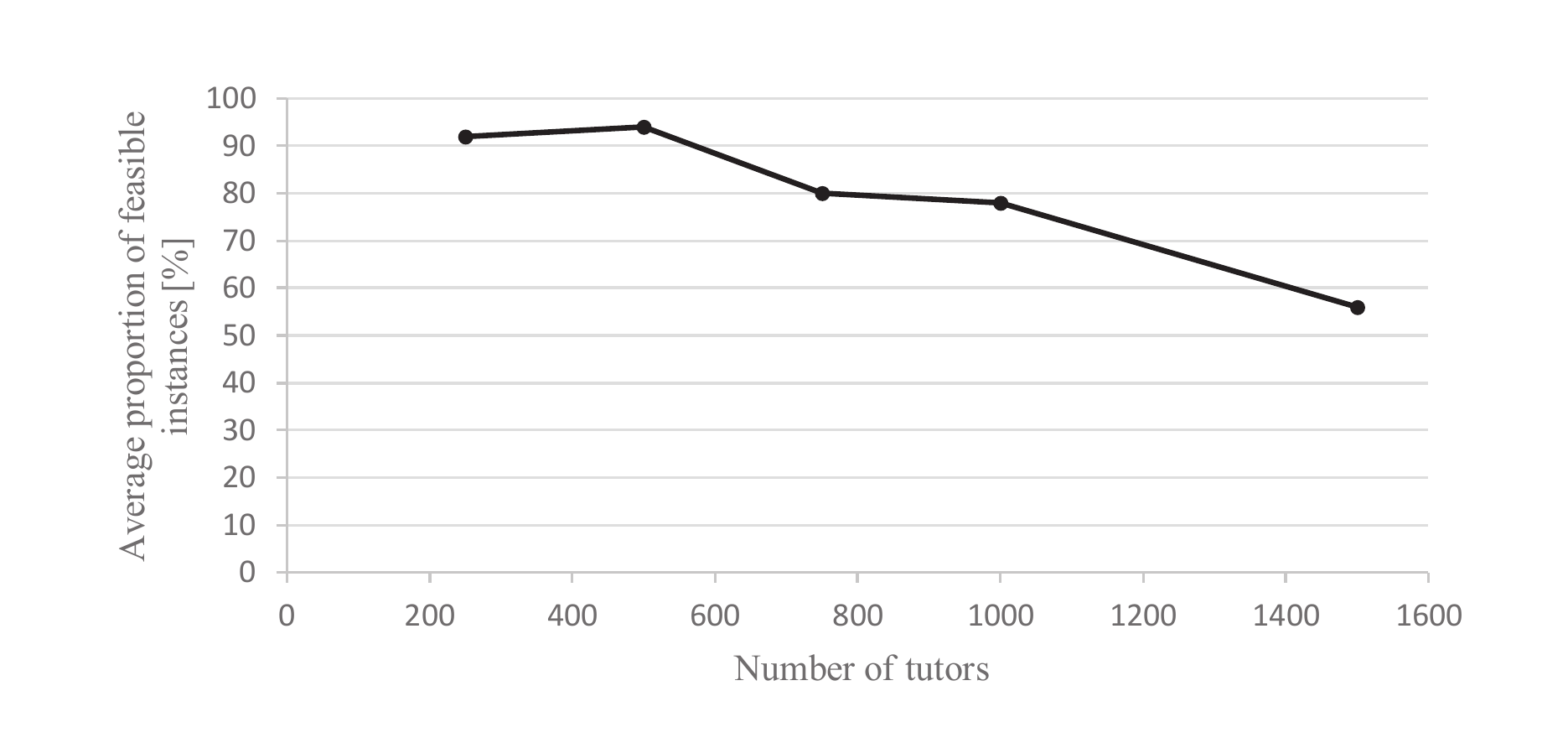}
    \caption{Percentage of feasible instances in each experiment.}
    \label{fig:optimalsol}
\end{figure}
We observe that as the number of tutors grow, the number of infeasible instances grow as well. Infeasibility is usually caused by the lower and upper limits on the number of tutoring hours per tutor, indicating that there can be an incentive to put constraints (\ref{eq3}) in the objective function, as done in other works such as \cite{AHA11} and \cite{GKDA15}. 

Table \ref{tab:stability} measures the stability of the model within each experiment through the standard deviation of the computation time required to solve the model and the standard deviation of the solution value. The coefficient of variation of the computational time (i.e., the standard deviation divided by the mean) is relatively small (around 0.1), which indicates that the time required to solve an instance is relatively similar for all the instances in the same group. In other words, the difficulty of our instances mainly comes from their large size, so any combinatorial technique reducing the number of constraints or variables (such as preprocessing) could significantly decrease the average computing time.

\begin{table}[t]
	\caption{Standard Deviations (SD) of model computing time and solution value for each set of experiments}
	\label{tab:stability}
	\centering
	\footnotesize
	\begin{tabular}{rrr rr}
		\toprule
		\multicolumn{1}{c}{N. tutors} & \multicolumn{1}{c}{N. courses} & \multicolumn{1}{c}{N. instances} & \multicolumn{1}{c}{SD of Model time [s]} & \multicolumn{1}{c}{SD of solution value} \\
		\cmidrule(lr){1-3} \cmidrule(lr){4-5}
		250 & 100 & 50 & 5.93 & 5.84 \\ 
		500 & 200 & 50 & 15.05 & 8.01 \\
		750 & 300 & 50 & 43.45 & 10.82 \\
		1000 & 400 & 50 & 44.39 & 10.85 \\
		1500 & 600 & 50 & 79.44 & 20.10 \\
		\bottomrule
	\end{tabular}
\end{table}

\subsection{Sensitivity analysis} \label{subsec:sensitivity_TAP}

We carried out a sensitivity analysis to evaluate model behavior and robustness with respect to small changes in the values of input parameters, observing the corresponding variations in the outputs of the model. 

First of all, we studied the differences in solution values by varying the maximum number of preferences expressed by the tutors. In the first experiment explained in Section \ref{subsec:genexp_TAP}, the number of preferences expressed by a tutor was an integer number randomly generated between 0 and 3 per semester, which corresponds to the maximum number of courses a tutor can be assigned to in each semester. Two additional experiments were performed, by generating the same instances as the first experiment, but only reducing and enlarging the maximum number of preferences per semester to two and four, respectively. The results we obtained are compared to the previous experiment as reported in Table \ref{tab:sensitivity1}. The solution value decreases on average by 3\% with a maximum of two preferences expressed by tutors and increases on average by 4\% with a maximum of four preferences. 

\begin{table}[tb]
	\caption{Sensitivity analysis on tutors' preferences.}
	\label{tab:sensitivity1}
	\footnotesize
	\centering
	\resizebox{\textwidth}{!}{
	\begin{tabular}{lrrrrrrrr}
		\toprule
		\multicolumn{1}{c}{Max N.} 		& \multicolumn{1}{c}{N. tutors} & \multicolumn{1}{c}{N. courses} & \multicolumn{1}{c}{N. instances} & \multicolumn{1}{c}{N. feasible}  & \multicolumn{1}{c}{N. optimal}  & \multicolumn{1}{c}{AVG model} & \multicolumn{1}{c}{AVG solution} \\
		\multicolumn{1}{c}{preferences}	& 		 & 	 	      &  		     &    \multicolumn{1}{c}{solutions} & \multicolumn{1}{c}{solutions} &  \multicolumn{1}{c}{time [s]} &   \multicolumn{1}{c}{value} \\
		\cmidrule(lr){1-1} \cmidrule(lr){2-4} \cmidrule(lr){5-8}
	\multirow{5}*{2} & 250 & 100 & 50 & 46 & 46 & 58.65 & 98.21 \\
		& 500 & 200 & 50 & 47 & 47 & 194.74 & 198.56 \\
		& 750 & 300 & 50 & 40 & 40 & 426.03 & 295.56 \\
		& 1000 & 400 & 50 & 39 & 39 & 690.50 & 396.62 \\
		& 1500 & 600 & 50 & 28 & 28 & 1610.54 & 598.88 \\
		\cmidrule(lr){1-1} \cmidrule(lr){2-4} \cmidrule(lr){5-8}
	\multirow{5}*{3} & 250 & 100 & 50 & 46 & 46 & 56.27 & 102.01 \\
		& 500 & 200 & 50 & 47 & 47 & 186.50 & 205.05 \\
		& 750 & 300 & 50 & 40 & 40 & 397.15 & 303.67 \\
		& 1000 & 400 & 50 & 39 & 39 & 663.83 & 408.36 \\
		& 1500 & 600 & 50 & 28 & 28 & 1474.68 & 614.30 \\
		\cmidrule(lr){1-1} \cmidrule(lr){2-4} \cmidrule(lr){5-8}
	\multirow{5}*{4} & 250 & 100 & 50 & 46 & 46 & 59.47 & 105.88 \\
		& 500 & 200 & 50 & 47 & 47 & 195.02 & 212.73 \\
		& 750 & 300 & 50 & 40 & 40 & 401.55 & 316.84 \\
		& 1000 & 400 & 50 & 39 & 39 & 749.59 & 425.05 \\
		& 1500 & 600 & 50 & 28 & 28 & 1516 & 636.59 \\
		\bottomrule
	\end{tabular}
	}
\end{table}

A second analysis was executed on the number of locations, in order to detect the importance of this parameter in the model, especially focusing on how it affects the model computation time. In all previous computational experiments, the number of locations was equal to two as we aimed at reproducing the real case presented in Section \ref{subsec:casestudy_TAP}. Two additional experiments were carried out by considering one and three locations, respectively. For the instances with 3 locations, 90\% of the workshops are in location 1, 5\% in location 2, and 5\% in location 3. Results are shown in Table \ref{tab:sensitivity2}, where they are compared with the results of the original experiment. While the solution value remains stable in the three experiments, the model time varies significantly. In particular, instances with only one location are solved with an average reduction of 10\% in the model computation time. When switching from two to three locations, we noticed an average increment of 33\% in the time to solve the model.

\begin{table}[t]
	\caption{Results obtained with the sensitivity analysis of number of locations.}
	\label{tab:sensitivity2}
	\footnotesize
	\centering
	\resizebox{\textwidth}{!}{
	\begin{tabular}{lrrrrrrrr}
		\toprule
		\multicolumn{1}{c}{N. locations} 		& \multicolumn{1}{c}{N. tutors} & \multicolumn{1}{c}{N. courses} & \multicolumn{1}{c}{N. instances} & \multicolumn{1}{c}{N. feasible}  & \multicolumn{1}{c}{N. optimal}  & \multicolumn{1}{c}{AVG model} & \multicolumn{1}{c}{AVG solution} \\
		\multicolumn{1}{c}{}	& 		 & 	 	      &  		     &    \multicolumn{1}{c}{solutions} & \multicolumn{1}{c}{solutions} &  \multicolumn{1}{c}{time [s]} &   \multicolumn{1}{c}{value} \\
		\cmidrule(lr){1-1} \cmidrule(lr){2-4} \cmidrule(lr){5-8}
	\multirow{5}*{1} & 250 & 100 & 50 & 44 & 44 & 48.03 & 100.25 \\
		& 500 & 200 & 50 & 43 & 43 & 174.41 & 201.67 \\
		& 750 & 300 & 50 & 41 & 41 & 357.16 & 305.00 \\
		& 1000 & 400 & 50 & 36 & 36 & 613.03 & 404.63 \\
		& 1500 & 600 & 50 & 31 & 31 & 1345.51 & 608.28 \\
		\cmidrule(lr){1-1} \cmidrule(lr){2-4} \cmidrule(lr){5-8}
	\multirow{5}*{2} & 250 & 100 & 50 & 46 & 46 & 56.27 & 102.01 \\
		& 500 & 200 & 50 & 47 & 47 & 186.50 & 205.05 \\
		& 750 & 300 & 50 & 40 & 40 & 397.15 & 303.67 \\
		& 1000 & 400 & 50 & 39 & 39 & 663.83 & 408.36 \\
		& 1500 & 600 & 50 & 28 & 28 & 1474.68 & 614.30 \\
		\cmidrule(lr){1-1} \cmidrule(lr){2-4} \cmidrule(lr){5-8}
	\multirow{5}*{3} & 250 & 100 & 50 & 46 & 46 & 81.33 & 101.95 \\
		& 500 & 200 & 50 & 47 & 47 & 251.55 & 204.90 \\
		& 750 & 300 & 50 & 40 & 40 & 519.84 & 303.48 \\
		& 1000 & 400 & 50 & 39 & 39 & 865.41 & 408.08 \\
		& 1500 & 600 & 50 & 28 & 28 & 1866.14 & 613.91 \\
		\bottomrule
	\end{tabular}
	}
\end{table}

\section{Conclusion}
\label{sec:conclusion_TAP}
This paper deals with the Tutor Allocation Problem, which aims at assigning tutors to workshops and is a key problem faced by many universities every year. This problem is important as tutors tend to give high-quality workshops when they are assigned to courses they wish to tutor. We proposed an Integer Linear Programming model that maximizes tutors' preferences for workshops, while satisfying a large set of constraints such as location, min/max number of hours, min/max number of courses, number of required tutors per workshop, scheduling, and incompatibility constraints.

To validate the model, three kinds of computational experiments were carried out. First, the 2019/2020 problem for the School of Mathematics at the University of Edinburgh was solved, and we showed that we could satisfy 60\% of the tutors' choices while respecting all constraints. This is a clear improvement from the hand-based assignment that only satisfied 31\% of the tutors' choices. Then, a more complete analysis based on randomly generated data showed that our model could potentially solve instances with up to 1500 tutors and 600 courses in reasonable time. We also observed that the difficulty of the problem mainly comes from the high number of variables and constraints, so that any combinatorial optimization technique aimed at reducing the model size could be beneficial. Finally, a sensitivity analysis was performed to evaluate the model behavior with respect to two important parameters of the model. We observed a variation around 4\% in the solution value adding or removing one unit  to the maximum number of preferences a tutor can express per semester. A different analysis on the number of locations considered in the problem showed a good stability in the solution value yet a significant variation in the model computation time. Indeed, a decrease of 10\% was measured removing one location while an average increase of 33\% was shown adding one location.

We conclude this work by reminding that, as the university is a dynamic environment, changes in the tutors' availabilities or in one or more of the course specifications could make the solution obsolete. Therefore, if a change occurred before the Integer Linear Programming solution is made public, we could simply rerun the model. If a change occurred instead after, readjustment would be more problematic as it is not an option to rerun the model and obtain a completely different solution if the tutors have been communicated a first assignment. One option could be to rerun our model with an alternative objective function with the aim of minimizing the differences between the new and the old assignments. However, the number of changes might still be significant because the model prevents any constraint violation. Thus, it is important that the administrative members continue to actively participate in the process, so that they are able to monitor any change required after the solution publication. Indeed, they have the knowledge about what constraints could possibility be violated in order to fix a solution without adding large perturbations, if the model was not able to do so. This represents an interesting future research direction.

Another interesting future research direction is the study of metaheuristic algorithms to be applied to the Tutor Allocation Problem as it has been done for similar problems in the literature. We mention in particular the work of \cite{LMR19} and \cite{HASSA19} who studied advanced heuristic techniques for the optimization of university timetabling and scheduling problems. These types of techniques could be beneficial to solve very large instances or problems characterized by additional complicating constraints.

\section*{Acknowledgments}
The authors want to thank Luke Caudrey and Stuart King for their support during this project. The help and the time they dedicated us were greatly appreciated. 

\section*{Fundings}
This research was supported by the Engineering and Physical Science Research Council through grant No. EP/P029825/1, and by University of Modena and Reggio Emilia through grant FAR 2018.

\bibliography{ms}
\bibliographystyle{apalike}

\end{document}